# A Novel Digital Watermarking Algorithm using Random Matrix Image

Mahimn Pandya
Smt. K.B.Parekh College of Computer Science, Bhavnagar University.

Hiren Joshi
Department of Computer Science, Gujarat University.

Ashish Jani
Department of Computer Science, Kadi Sarva Vishwavidhyalay

## ABSTRACT

The availability of bandwidth for internet access is sufficient enough to communicate digital assets. These digital assets are subjected to various types of threats. [19] As a result of this, protection mechanism required for the protection of digital assets is of priority in research. The threat of current focus is unauthorized copying of digital assets which give boost to piracy. This under the copyright act is illegal and a robust mechanism is required to curb this kind of unauthorized copy. To safeguard the copyright digital assets, a robust digital watermarking technique is needed. The existing digital watermarking techniques protect digital assets by embedding a digital watermark into a host digital image. This embedding does induce slight distortion in the host image but the distortion is usually too small to be noticed. At the same time the embedded watermark must be robust enough to with stand deliberate attacks. There are various techniques of digital watermarking but researchers are making constant efforts to increase the robustness of the watermark image. The layered approach of watermarking based on Huffman coding [5] can soon increase the robustness of digital watermark.[11] Ultimately, increasing the security of copyright of protection. The proposed work is in similar direction where in RMI (Random Matrix Image) is used in place of Huffman coding. This innovative algorithm has considerably increased the robustness in digital watermark while also enhancing security of production.

## 1. INTRODUCTION

The Internet usage to transfer electronic assets has involved a technique that is able to protect the copyright of published medias into a certainty. The easy distribution of these documents through the web may violate protection laws against unauthorized copies and make fidelity questionable. Digital watermarking has been projected as a solution against these practices. Digital watermark is an authenticating technique of digital data with secret information that can be extracted to the receptor. The image in which this data is inserted is called 'cover image' or 'host image'. The watermarking process has to be resistant against possible attacks, keeping the content of the watermark readable in order to be recognized when extracted. Features like robustness and fidelity are essentials of a watermarking system however the size of the embedded information has to be considered since data becomes less robust as its size increases. Therefore a trade-off of these features must be considered. [1, 6, 10, 16 and 17]

## 2. PRINCIPLE OF DIGITAL WATERMARKING

A digital watermarking process have three phases, first embedding, second attack and third detection. In Embedding, an algorithm accepts the host image and the watermark image or data to be embedded and produces a watermarked image. The watermarked image is transmitted or stored, usually transmitted to another person. If this person makes a modification, this is called an attack.[12, 13, 14, 15, 20, 21] There are various kinds of attacks like copy, removal, mosaic etc. Watermark detection is an algorithm which is used to find the attacked data to attempt to extract the watermark from it. If the watermarked image is not modified during transmission, then the watermark is still present and it can be extracted. If the watermarked image is copied, then the information is also carried in the copy. The embedding takes place by manipulating the content of the digital data, which means the information is not embedded in the frame around the data, but it is carried with the watermarked image itself. [2] Figure 1 and Figure 2 shows the flowchart of watermarking process.

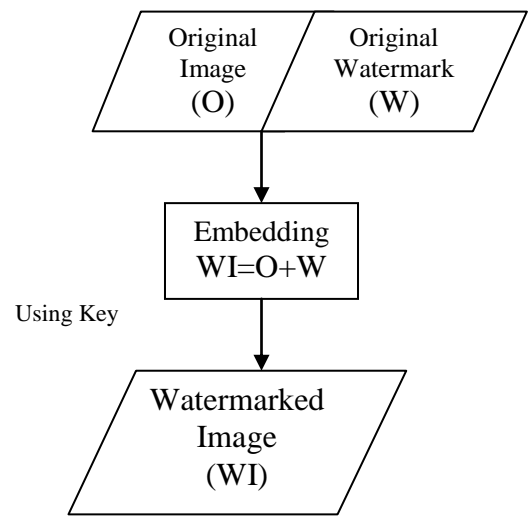

**Fig.1 Flowchart of Embedding of Watermark**

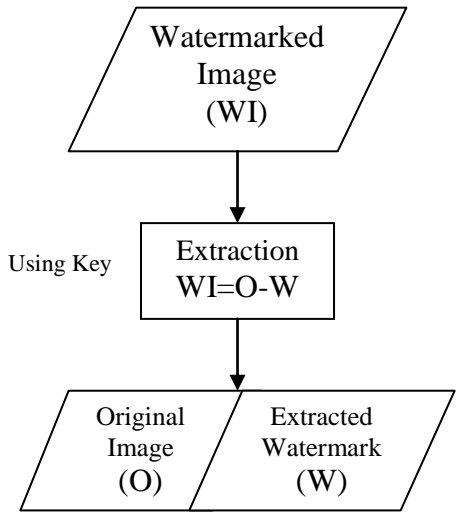

**Fig.2 Flowchart of Extraction of Watermark**

The original image and the watermark are embedded using one of the watermarking schemes that are available.[8,9] The watermarked image is processed through a detector in which generally a reverse process to that used during the embedding phase is applied to retrieve the watermark. The different watermarking algorithms differ in the way in which it embeds the watermark on to the cover object. A secret key is used during the embedding and the extraction process in order to prevent illegal access to the watermark [2, 22, and 7]. This paper deals with the new watermarking technique which helps to protect digital image based on RMI (Random Matrix Image).

## 3. PROPOSED WORK

In this work we are embedding Random Matrix Image as a watermark. We generate a new watermark image for each and every new image for watermarking. The proposed algorithms are used for watermark embedment and extraction. in watermark embedding we used a RMI to generate unique watermark and this watermark is being used to embed to an digital image. The exact reverse process is used to extract watermark for an image. For extraction we require RMI or an original image.

### 3.1 Overview of random matrix image.

This is an auto generated Image based on Random Matrix Image generated in SCILAB using random function [3]. In SCILAB Random Matrix can be generated using random function having randomised number from given range. In simulation we can also generate real number. For example we want to generate a random matrix of 8 x 8 from 0 to 10 numbers. Matrix may cover any number from 0 to 10 like shown in figure 3 (b).

### 3.2 Watermark embedding algorithm

Step1: Read the original image.

Step2: Generate RMI (in range of 0 to 10) which is to be embedded. (Secret Key Matrix)

Step3: Add this Generated Image and Original Image in matrix addition form.

Step4: Now generate image from matrix form.

Step5: The output image is a watermarked image.

### 3.3 Watermark extraction algorithm

Step1: Read the watermarked image.

Step2: Read matrix (a secret key) which is sent with image.

Step3: Subtract Matrix from watermarked Image in matrix subtraction form.

Step4: Now generate two different images from theses matrices form.

Step5: The output images are Original image and watermarked image.

## 4. IMPLIMENTATION & RESULTS

SCILAB simulations are performed by using 256×256 pixel gray level "Lena" image and 256 × 256 pixel watermark (RMI)[4]. Figure 3. (a) and (b) Shows greyscale image of 256X256 pixel and RMI Watermark respectively. Figure 4 (a) and (b) show 8X8 pixel matrix of Figure 3 (a) and (b) images respectively. Figure 5 (a) and (b) shows output image and 8x8 matrix of .watermarked Lena image.

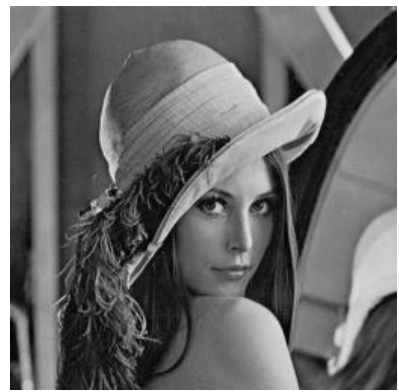

**(a)**

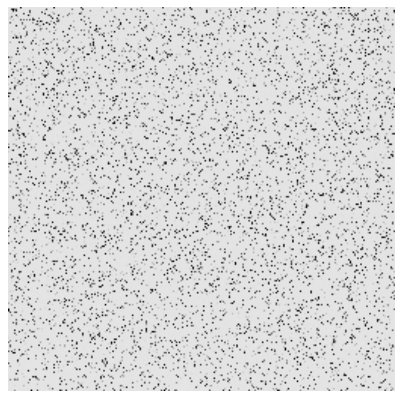

**(b)**

**Fig. 3 (a) Original Lena Image (b) RMI watermark**

| 195 | 195 | 196 | 197 | 197 | 198 | 199 | 199 |
| --- | --- | --- | --- | --- | --- | --- | --- |
| 196 | 196 | 196 | 197 | 197 | 197 | 198 | 198 |
| 197 | 197 | 197 | 197 | 196 | 196 | 196 | 196 |
| 199 | 198 | 198 | 197 | 196 | 195 | 194 | 194 |
| 199 | 198 | 197 | 196 | 195 | 194 | 193 | 193 |
| 198 | 198 | 197 | 196 | 195 | 194 | 193 | 193 |
| 197 | 196 | 196 | 195 | 195 | 194 | 194 | 194 |
| 196 | 196 | 195 | 195 | 195 | 194 | 194 | 194 |

**(a)**

| 2 | 2 | 9 | 8 | 2 | 6 | 3 | 6 |
| --- | --- | --- | --- | --- | --- | --- | --- |
| 7 | 2 | 0 | 0 | 0 | 0 | 8 | 3 |
| 9 | 1 | 3 | 0 | 4 | 0 | 1 | 1 |
| 2 | 6 | 3 | 0 | 7 | 2 | 4 | 2 |
| 9 | 3 | 0 | 6 | 3 | 6 | 7 | 2 |
| 5 | 5 | 3 | 7 | 7 | 0 | 5 | 2 |
| 5 | 1 | 0 | 7 | 5 | 8 | 0 | 5 |
| 2 | 0 | 5 | 5 | 9 | 7 | 6 | 6 |

**(b)**
**Fig.4 (a) 8x8 matrix of Lena (b) 8x8 matrix of RMI**

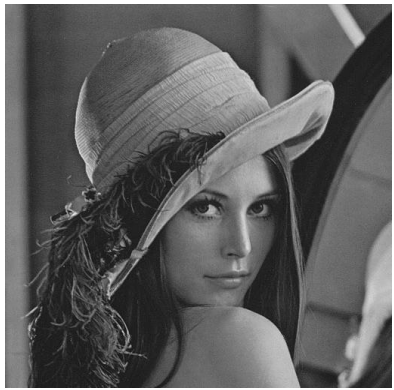

**(a)**

| 197 | 197 | 205 | 205 | 199 | 204 | 202 | 205 |
| --- | --- | --- | --- | --- | --- | --- | --- |
| 203 | 198 | 196 | 197 | 197 | 197 | 206 | 201 |
| 206 | 198 | 200 | 197 | 200 | 196 | 197 | 197 |
| 201 | 204 | 201 | 197 | 203 | 197 | 198 | 196 |
| 208 | 201 | 197 | 202 | 198 | 200 | 200 | 195 |
| 203 | 203 | 200 | 203 | 202 | 194 | 198 | 195 |
| 202 | 197 | 196 | 202 | 200 | 202 | 194 | 199 |
| 198 | 196 | 200 | 200 | 204 | 201 | 200 | 200 |

**(b)**
**Fig.5 (a) Watermarked Image (b) 8x8 matrix of Fig.5(a)**

In figure 5 (a) shows that watermarked Lena image and (b) shows 8x8 matrix of watermarked the image. Figure. 5(b) surely shows the addition of two matrices shown in image Figure. 4 (a) and (b). The change in pixel vales that shows embedment of RMI to Lena image.

## 5. CONCLUSION

A novel method of digital watermarking based on embedding matrix as a watermark is presented in this paper. The experiment part has used random matrix as a watermark to prevent an attacker for easy attack on the water mark image. The reason for this is that each image usually has different matrix form with range from 0 to 10. The noticeable part of this work is the use of RMI transition to authenticated user of the image. Without having RMI, none other than authenticated user can detect and extract watermark from watermarked image. The limitation of the current work is that the stated technique of digital water marking is used on grey scale image and not on the coloured image. The experiment performed on greyscale image having less than 245 greyscale pixel value. The extension of this work can cover the use of stated technique of digital water marking on colour images.